\documentclass[twocolumn,draft,prd,superscriptaddress,showpacs,floatfix%
,preprintnumbers,nofootinbib]{revtex4}

\usepackage[final]{epsfig}
\usepackage{bm}
\usepackage[usenames]{color}

\begin{document}

\title{High-resolution supernova neutrino spectra represented by a simple fit}

\author{Irene Tamborra}
\affiliation{Max-Planck-Institut f\"ur Physik
(Werner-Heisenberg-Institut), F\"ohringer Ring 6, 80805 M\"unchen,
Germany}

\author{Bernhard M\"uller}
\affiliation{Max-Planck-Institut f\"ur Astrophysik,
Karl-Schwarzschild-Str.~1, 85748 Garching, Germany}

\author{Lorenz H\"udepohl}
\affiliation{Max-Planck-Institut f\"ur Astrophysik,
Karl-Schwarzschild-Str.~1, 85748 Garching, Germany}

\author{Hans-Thomas Janka}
\affiliation{Max-Planck-Institut f\"ur Astrophysik,
Karl-Schwarzschild-Str.~1, 85748 Garching, Germany}

\author{Georg Raffelt}
\affiliation{Max-Planck-Institut f\"ur Physik (Werner-Heisenberg-Institut),
F\"ohringer Ring 6, 80805 M\"unchen, Germany}

\date{\today}

\preprint{MPP-2012-149}

\begin{abstract}
To study the capabilities of supernova neutrino detectors, the
instantaneous spectra are often represented by a quasi-thermal
distribution of the form $f_\nu(E)\propto E^\alpha
e^{-(\alpha+1)E/E_{\rm av}}$ where $E_{\rm av}$ is the average energy
and $\alpha$ a numerical parameter. Based on a spherically symmetric
supernova model with full Boltzmann neutrino transport we have, at a few
representative post-bounce times, re-converged the models with vastly
increased energy resolution to test the fit quality. For our examples,
the spectra are well represented by such a fit in the sense that the
counting rates for a broad range of target nuclei, sensitive to
different parts of the spectrum, are reproduced very well. Therefore,
the  mean energy and root-mean-square energy of
numerical spectra hold enough information to provide the
  correct $\alpha$ and to forecast the response of multi-channel
supernova neutrino detection.
\end{abstract}


\maketitle

\section{Introduction}

The neutrino signal from the next nearby core-collapse supernova
(SN) remains the most coveted target for low-energy neutrino
astronomy. Galactic SNe are rare, perhaps a few per century, and
such an observation will be a once-in-a-lifetime opportunity to look
deeply inside a collapsing star and learn about its astrophysical
workings as well as about neutrino properties. Several detectors
worldwide are in operation that will register a high-statistics
signal while others are in preparation or under
discussion~\cite{Scholberg:2012id}.

In order to assess the physics potential of various detectors or
detection principles one needs the expected flavor-dependent SN
neutrino flux spectra. In the absence of self-consistent
three-dimensional core-collapse simulations there is no standard SN
neutrino flux model and moreover, the flavor-dependent flux spectra
depend on the properties of the incompletely known neutron-star
equation of state, on the properties of the collapsing star, 
notably its mass,
and, of course, on time for any given case. In this situation
parametric studies, taking account of the plausible range of
predictions, are the preferred course of action.

We here address one particular aspect of such studies, i.e.\ the
plausible spectral shape of SN neutrino fluxes. On a rough level of
approximation, the spectra follow a thermal distribution that can be
described in terms of an effective temperature. In detail the
spectra formation is a complicated process, different energy groups
emerging from different depths in the proto-neutron star
atmosphere~\cite{Raffelt:2001kv, Keil:2002in}, and a more refined
description is necessary for a more detailed understanding.

The next level of sophistication is to describe the non-equilibrium
spectra by a three-parameter fit that allows for deviations from a
strictly thermal spectrum~\cite{Janka:1989}.  One particularly simple
realization are spectra where  the 
energy-dependent neutrino number flux for each flavor has the
  form~\cite{Keil:2002in}
\begin{equation}\label{eq:alphafit}
f_\nu(E)\propto E^\alpha\,e^{-(\alpha+1) E/E_{\rm av}}\,.
\end{equation}
Here, $E_{\rm av}$ is the average energy and $\alpha$ a numerical
parameter describing the amount of spectral pinching;
the value $\alpha=2$ corresponds to a Maxwell-Boltzmann spectrum,
 and $\alpha=2.30$ to a Fermi-Dirac distribution with
  zero chemical potential. In general the neutrino spectra are fitted
  by $2 \le \alpha \le 4$~\cite{Keil:2002in} with higher $\alpha > 2$
  indicating stronger pinching and $\alpha <2$ meaning anti-pinching
  relative to a Maxwell-Boltzmann distribution. The three parameters
$E_{\rm av}$, $\alpha$, and the overall normalization can be
determined, for example, if a numerical SN simulation provides the
energy flux (luminosity) $L_\nu$ in some flavor, the average energy
$E_{\rm av}=\langle E_\nu\rangle$ and some other energy moment, for
example $\langle E_\nu^2\rangle$ or sometimes $\langle
E_\nu^3\rangle$.

While spectra of the form of Eq.~(\ref{eq:alphafit}) certainly provide
a reasonable overall representation, it is not obvious how well the
spectral tails are reproduced.   For studying detector
  responses with target nuclei with a significant energy threshold, it
  is imperative that the accuracy of the $\alpha$-fit in the
  high-energy tail be checked against solutions of the neutrino
  transport equations. This is especially relevant for example for
  lead in the Halo detector~\cite{Vaananen:2011bf}, argon in future large-scale liquid argon
  detectors~\cite{GilBotella:2004bv}, or subdominant detection channels on oxygen in water
  Cherenkov detectors~\cite{Langanke:1995he} or on carbon in liquid scintillator detectors~\cite{Wurm:2011zn}.

Modern numerical SN codes treat neutrino transport with Boltzmann
solvers that are expected to produce physically accurate spectra. On
the other hand, neutrino transport is the most CPU-time consuming
aspect of SN simulations so that in practice the energy resolution is
limited. For example, in typical simulations of the Garching group,
17--21 energy bins are used, corresponding to an energy
  resolution $\Delta \epsilon / \epsilon \sim 0.3$.  Whether such a
  (seemingly) coarse zoning is adequate for modeling the high-energy
  tail of the spectrum and hence for judging the quality of
  $\alpha$-fits needs to be determined at least once by a resolution
  study. Spectra from standard resolution (SR) neutrino transport
  simulations are therefore of little value for assessing the quality of
  $\alpha$-fits in the spectral tail, as are Monte Carlo studies
  \cite{Janka:1989,Keil:2002in}, which suffer from limited
  sampling at high energies.

In this paper, we investigate whether $\alpha$-fits and
SR spectra from SN simulations are sufficiently accurate
to predict detection rates that are sensitive to the high-energy tail.
To this end, we
study the spectral shape and its impact on the relative
detector response of different materials
 using \emph{high-resolution} (HR) spectra computed for
a few representative post-bounce times of a representative spherically
 symmetric simulation.  These spectra have
been obtained by re-converging models at a given post-bounce time
  after refining the energy binning. With the HR spectra
  we can then test how well a global fit of the form of
  Eq.~(\ref{eq:alphafit}) reproduces the relative neutrino counting
  rates of different materials that probe rather different parts of
  the spectrum.
  
  We always ignore the effect of neutrino flavor conversions. The
ordinary Mikheyev-Smirnov-Wolfenstein (MSW) effect~\cite{wolf,Wolfenstein:1977ue} or self-induced flavor conversions~\cite{Duan:2010bg} 
strongly modify the flavor-dependent spectra seen by a detector on Earth~\cite{Fogli:2008fj}. For
example, the detectable $\bar\nu_e$ flux will be a certain linear
combination of the $\bar\nu_e$ and $\bar\nu_x$ spectra emitted at
the source. These detection spectra need not be represented by a
simple fit, in particular if self-induced conversions cause features
such as ``spectral splits.'' Determining the detection spectra in
terms of the source spectra is the task of oscillation studies,
usually performed by post-processing the fluxes provided by
numerical models. Justifying simple representations of the source
spectra for the purpose of flavor conversion studies is one prime
motivation for our work.

To this end we describe in Sec.~\ref{sec:SNmodel} our numerical SN
model and in Sec.~\ref{sec:nuclei} the target materials. A
comparison between counting rates based on the numerical spectra and
the fitted ones is performed in Sec.~\ref{sec:comparison} before
concluding in Sec.~\ref{sec:conclusions}.

\begin{figure*}
\includegraphics[width=1.\linewidth]{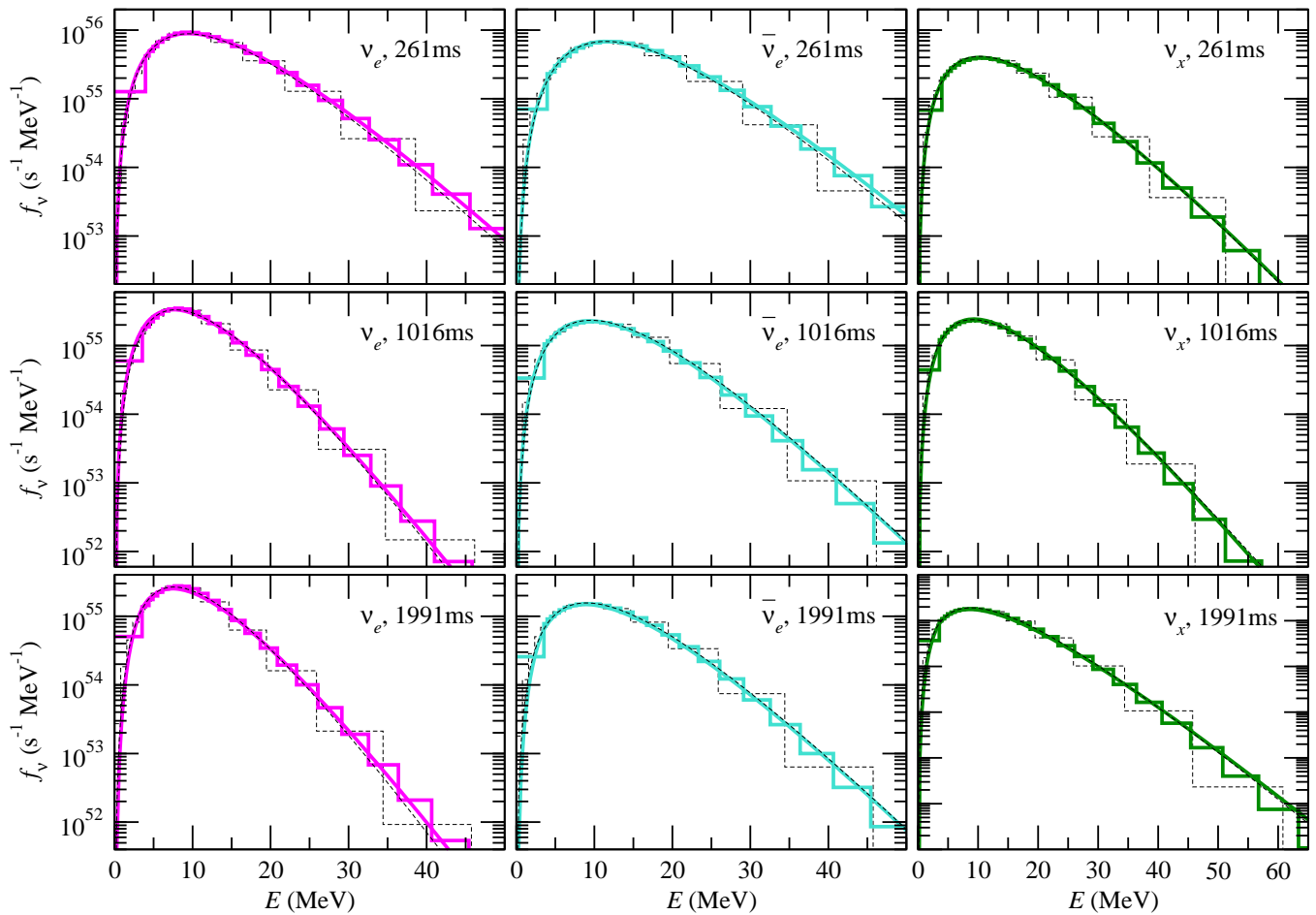}
\caption{\label{fig:log_spectra}Spectra for species $\nu_e$ (left
  column), $\bar{\nu}_e$ (middle column) and $\nu_x$ (right column)
  for post-bounce times of $261$ (top row), $1016$ (middle row),
  and $1991~\mathrm{ms}$ (bottom
  row). The data from the \textsc{Prometheus-Vertex} spectra are
  shown as step functions, and the continuous curves are quasi-thermal
  fits according to Eq.~(\ref{eq:alphafit}). Thick magenta, turquoise, and green lines are used for the
different neutrino species ($\nu_e$, $\bar{\nu}_e$ and $\nu_x$, respectively) in the HR case, while 
the SR spectra and fits are shown as thin dashed black lines.}
\end{figure*}
\begin{figure}
\includegraphics[width=1.\linewidth]{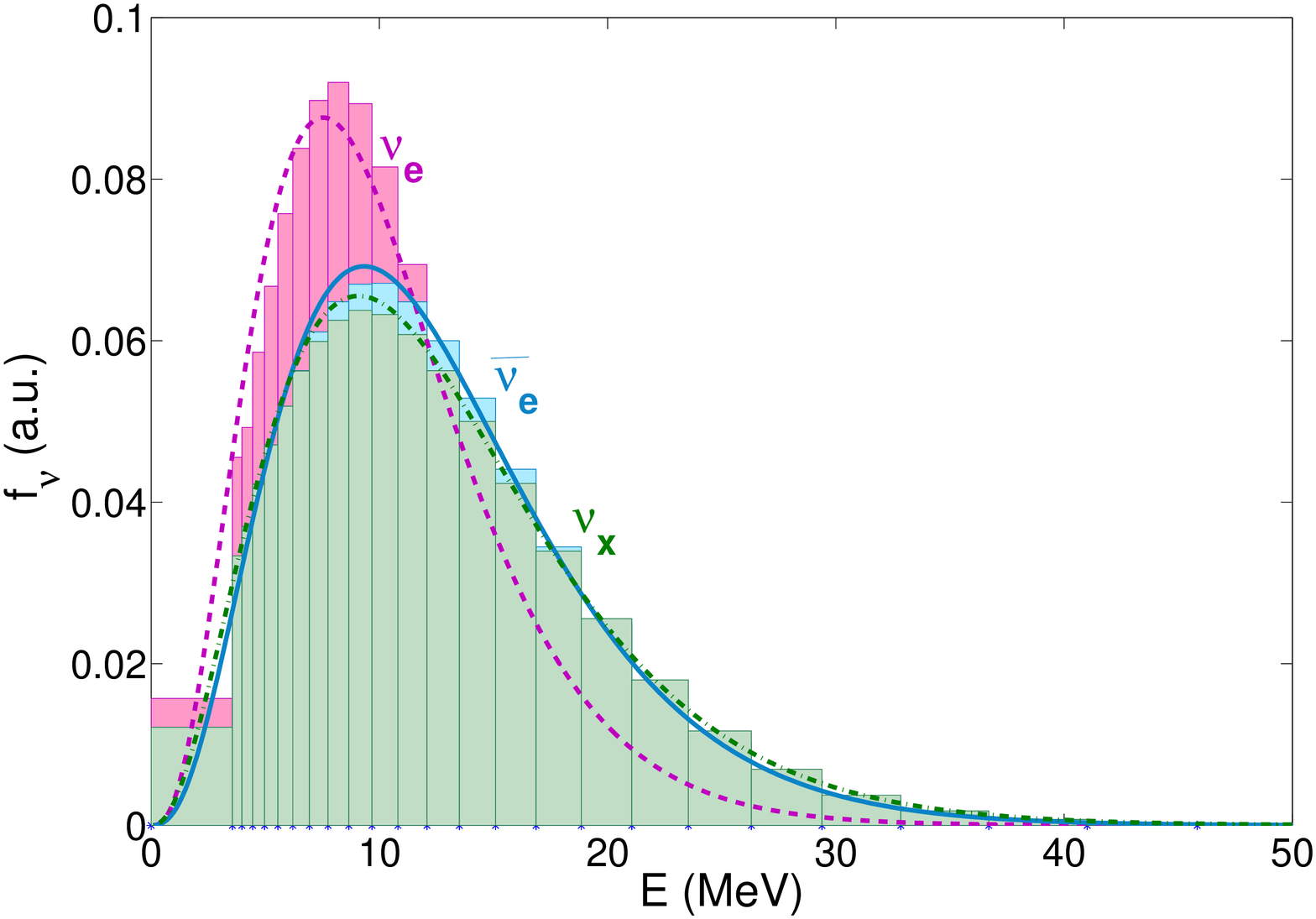}
\caption{\label{fig:spectra}Spectra for species
$\nu_e$, $\bar{\nu}_e$ and $\nu_x$. Histograms are the \textsc{Prometheus-Vertex} HR spectra
of our model at time 1016~ms.  The curves are quasi-thermal fits
according to Eq.~(\ref{eq:alphafit}): dashed for $\nu_e$, continuous for $\bar{\nu}_e$, and dot-dashed for $\nu_x$.}
\end{figure}

\section{Numerical supernova model}                \label{sec:SNmodel}

We have performed simulations of the $15 M_\odot$ progenitor of
Woosley and Weaver~\cite{Woosley:1995} with the neutrino
hydrodynamics code \textsc{Prometheus-Vertex} \cite{Rampp:2002},
which solves the 0th and 1st neutrino moment equations with a
variable Eddington factor closure that is provided by the formal
solution of a simplified (model) Boltzmann equation. The explosion of
the spherically symmetric model was artificially initiated $500 \
\mathrm{ms}$ after bounce. For the dynamical evolution of the model
including the feedback into the hydrodynamics, we used 21 energy bins
with a typical resolution $\delta \epsilon / \epsilon$ (ratio of zone
width to zone mean energy) of $\sim 0.29$. We refer to this setup as
the SR case.

In order to accurately predict the shape of the neutrino spectra,
particularly in the high-energy tail, we computed high-resolution (HR)
stationary solutions of the neutrino transport equations for a
representative time during the accretion phase ($261 \ \mathrm{ms}$
after bounce) and for the proto-neutron star cooling phase ($1016
\ \mathrm{ms}$ and $1991 \ \mathrm{ms}$) using the matter background
from the dynamical simulation.   Different from the dynamical run,
  the original treatment of \cite{Rampp:2002} for the Doppler and redshift terms is used
instead of that of \cite{Mueller:2010}, and the threshold values for
the monochromatic neutrino energy density below which a spectral
extrapolation is performed has been reduced by a factor of 100.  We
also transition continuously from the comoving frame to the lab frame
once neutrino interactions cease in order to eliminate artifacts due
to the velocity jump in the shock.  These numerical changes accelerate
convergence and help to obtain smoother and more accurate spectra.

The spacing of the logarithmic energy grid was much finer with 42 bins
and $\delta \epsilon / \epsilon = 0.11$  in the HR case.  For
better comparison, the stationary solutions for the SR case were
re-computed with precisely the same
numerics. Figure~\ref{fig:log_spectra} shows the obtained HR spectra
for the species $\nu_e$, $\bar{\nu}_e$,  and $\nu_x$
 (with $\nu_x$ meaning a \emph{single} kind of
$\nu_\mu$, $\bar\nu_\mu$, $\nu_\tau $, and $\bar\nu_\tau$).

\begin{table*}
\caption{Fit parameters $\alpha$, luminosities $L_\nu$, and neutrino
mean energies $\langle E_\nu \rangle$ for different neutrino species
and post-bounce times for HR and SR spectra. \label{tab:table1} }
\begin{ruledtabular}
\begin{tabular}{llllllll}
$t$   & species       &  $\alpha_\mathrm{HR}$ & $\alpha_\mathrm{SR}$ &  $L_{\nu,\mathrm{HR}}$     & $L_{\nu,\mathrm{SR}}$      &  $\langle E_\nu \rangle_\mathrm{HR}$ & $\langle E_\nu \rangle_\mathrm{SR}$ \\
$[\mathrm{ms}]$  &               &                       &                      & $[10^{52}\,\mathrm{erg\,s}^{-1}]$ & $[10^{52}\,\mathrm{erg\,s}^{-1}]$ &  $[\mathrm{MeV}]$  & $[\mathrm{MeV}]$ \\
 \hline
261               & $\nu_e$          &  2.65              & 2.73               & 2.80          & 2.68         &  13.05                         & 12.87 \\
261               & $\bar{\nu}_e$    &  3.13              & 3.23               & 2.79          & 2.71         &  15.23                         & 15.05 \\
261               & $\nu_x$          &  2.42              & 2.48               & 1.58          & 1.63         &  14.48                         & 14.48 \\
\hline
1016              & $\nu_e$          &  2.90              & 3.14               & 0.62         & 0.62        &  10.14                         & 10.28 \\
1016              & $\bar{\nu}_e$    &  2.78              & 2.88               & 0.67         & 0.70        &  12.69                         & 12.87 \\
1016              & $\nu_x$          &  2.39              & 2.52               & 0.75         & 0.77        &  12.89                         & 13.00 \\
\hline
1991              & $\nu_e$          &  2.92              & 3.17               & 0.48         & 0.47        & 10.01                          & 10.01 \\
1991              & $\bar{\nu}_e$    &  2.61              & 2.67               & 0.44         & 0.44        & 12.28                          & 12.36 \\
1991              & $\nu_x$          &  2.34              & 2.49               & 0.54         & 0.55        & 12.31                          & 12.39 \\
\end{tabular}
\end{ruledtabular}
\end{table*}

In order to compare the numerical spectra (histograms) with a
quasi-thermal fit of the form of Eq.~(\ref{eq:alphafit}), we determine
the parameter $\alpha$ in terms of the energy moments of the
distribution. One way of fixing $\alpha$ is to use the energy moment
of order $k$ and notice that
\begin{equation}
\frac{\langle E^k_\nu \rangle}{\langle E^{k-1}_\nu \rangle}
=\frac{k+\alpha}{1+\alpha}\, \langle E_\nu \rangle\,.
\end{equation}
We have checked that for our cases the implied value of $\alpha$
does not strongly depend on $k$, which already implies that this
type of fit is a reasonable representation. In the following we will
use $k=2$, i.e., the relation
\begin{equation}
\frac{\langle E^2_\nu \rangle}{\langle E_\nu\rangle^2}=\frac{2+\alpha}{1+\alpha}\,.
\label{eq:ratio}
\end{equation}
We show examples of the fits as continuous colored curves alongside
the HR spectra (step functions) on a logarithmic and
on a linear scale in Figs.~\ref{fig:log_spectra} and \ref{fig:spectra}, respectively.
For comparison we also show the SR spectra and
their correspondent fits as thin dashed black lines in Fig.~\ref{fig:log_spectra}, and  find
that visually the fits work very well in both cases.   Values for the fit
parameter $\alpha$, the luminosity $L_\nu$, and the neutrino mean
energy $\langle E_\nu \rangle$ for the three snapshots and for the
different neutrino species $\nu_e$, $\bar{\nu}_e$, and $\nu_x$ are
listed in Table~\ref{tab:table1} both for the HR and the SR case. We
note that despite the lower energy resolution, the SR spectra from the
dynamical simulation typically give very similar luminosities and mean
energies, which are consistent with the HR case within $\sim 5\%$ or
less. The fit parameter $\alpha$ agrees similarly well except for
$\nu_e$ in the post-explosion phase with a deviation of up to
$9\%$. However, this is due to a very steep dependence of $\alpha$ on
the second energy moment, $\langle E_\nu^2 \rangle$, when the energy
ratio in Eq.~(\ref{eq:ratio}) is close to unity, and does not affect
the fit curves appreciably.

\section{Detection cross sections}                  \label{sec:nuclei}

In existing detectors, SN neutrinos are primarily measured in the
$\bar\nu_e$ channel by virtue of the inverse beta decay (IBD)
reaction $\bar{\nu}_e+p \rightarrow n+e^+$. The cross section varies
roughly as $E_\nu^2$ so that the overall event rate is well
reproduced, if the SN neutrino spectrum is approximated by a
quasi-thermal spectrum where the second energy moment is used to
determine $\alpha$.

However, other detection channels are also of interest. One is
elastic scattering on electrons where the cross section varies
roughly linearly with $E_\nu$. Of greater interest  for our purpose of testing 
the goodness of the $\alpha$-fit are
reactions with a higher energy threshold or steeper energy variation
than IBD. One example is the recently built HALO detector in SNOLAB
where 79 tons of lead are used as target material. Other examples
are reactions on argon in future liquid argon detectors that are
primarily considered in the context of long-baseline neutrino
oscillation studies and reactions on oxygen in water Cherenkov
detectors and on carbon in liquid scintillator detectors.

\begin{figure}
\includegraphics[width=1.\linewidth]{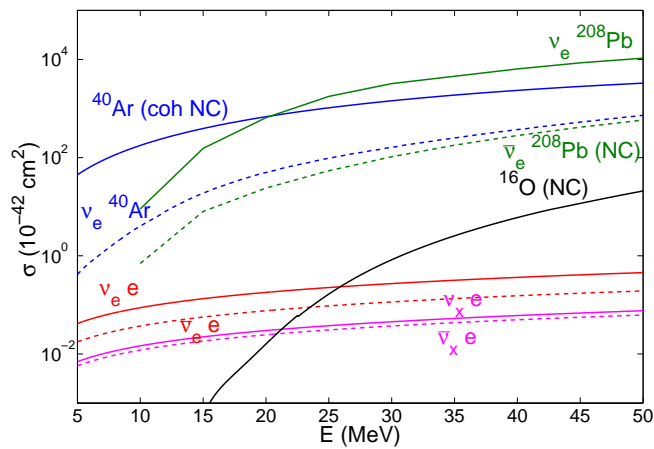}
\caption{\label{fig:sigma} Neutrino cross sections on some
representative target nuclei (see text for details).}
\end{figure}

In our study we consider several selected reactions that span a
broad range of spectral responses to neutrino fluxes and show the
energy-dependent cross sections in Fig.~\ref{fig:sigma}. The
reaction sensitive to the lowest range of energies is elastic
scattering on electrons via both charged-current (CC) and neutral-current (NC) for $\nu_e$ and
$\bar{\nu}_e$ and via NC interaction for $\nu_x$
\cite{Marciano:2003eq}.

In nuclei, CC interactions proceed via $\nu_e + (N,Z) \rightarrow
(N-1,Z+1) + e^-$ and $\bar{\nu}_e + (N,Z) \rightarrow (N+1,Z-1)+e^+$,
respectively. The $\bar{\nu}_e$ interaction is typically suppressed
at a given energy with respect to the $\nu_e$ interaction due to
Pauli blocking. Neutral current interactions may also produce observable signals
via ejected nucleons or de-excitation photons. As one example we
consider NC interactions with $^{16}{\mathrm
O}$~\cite{Langanke:1995he, Kolbe:2002gk, Haxton:1988mw,
Haxton:1990ks}, which is relevant for water-based detectors.

Interactions with heavier nuclei, such as lead, may yield quite high
rates of both CC and NC interactions. Observable signatures include
leptons and ejected nucleons. Single and  multiple neutron ejections
are possible. The relevant interactions for lead-based detectors are
$\nu_e +{}^{A}{\rm Pb} \rightarrow e^- +{}^{A}{\rm Bi}^*$ and $\nu_x
+{}^{A}{\rm Pb} \rightarrow \nu_x +{}^{A}{\rm Pb}^*$. For both CC and
NC cases, the resulting nuclei de-excite via neutron emission.
Antineutrino CC interactions are strongly suppressed. Although
natural lead contains isotopes other than $^{208}{\rm Pb}$, the
neutron cross section for $A=208$ should be similar to 
 other components. For the Pb cross sections see Refs.~\cite{Fuller:1998kb,
Kolbe:2000np, Toivanen:2001re, Samana:2008pt}.

Liquid argon detectors will have excellent sensitivity to $\nu_e$ via
the CC interaction $\nu_e +{}^{40}{\rm Ar} \rightarrow e^-
+{}^{40}{\rm K}^{*}$. This is an interaction for which the
de-excitation $\gamma$s from $^{40}{\rm K}^{*}$ can be observed. The
reaction $\bar{\nu}_e+{}^{40}{\rm Ar} \rightarrow e^++{}^{40}{\rm
  Cl}^{*}$ will also occur and can be tagged via the pattern of
$\gamma$s. NC excitations are possible, although little information is
available in the literature  because of the low-energy
  nuclear recoils that make it practically difficult to
  detect. Energy thresholds as low as few MeV may be
possible.  Although  the chances for
    detecting such a reaction are rather low, we consider NC coherent
  scattering anyway as an example of a reaction scanning the low-energy
  region of the neutrino spectra. The cross sections are reported in
Refs.~\cite{Raghavan:1986hv, GilBotella:2003sz, Bueno:2003ei,
  GilBotella:2004bv, SajjadAthar:2004yf}.

For the different channels that we consider, we use the cross
sections as in~\cite{Scholberg:2012id}. They have
been used in the SNOWGLOBES software package~\cite{snowglobes}.

\section{Counting rates}                        \label{sec:comparison}
\begin{figure}
\vspace{-0.2cm}
\includegraphics[width=1.\linewidth]{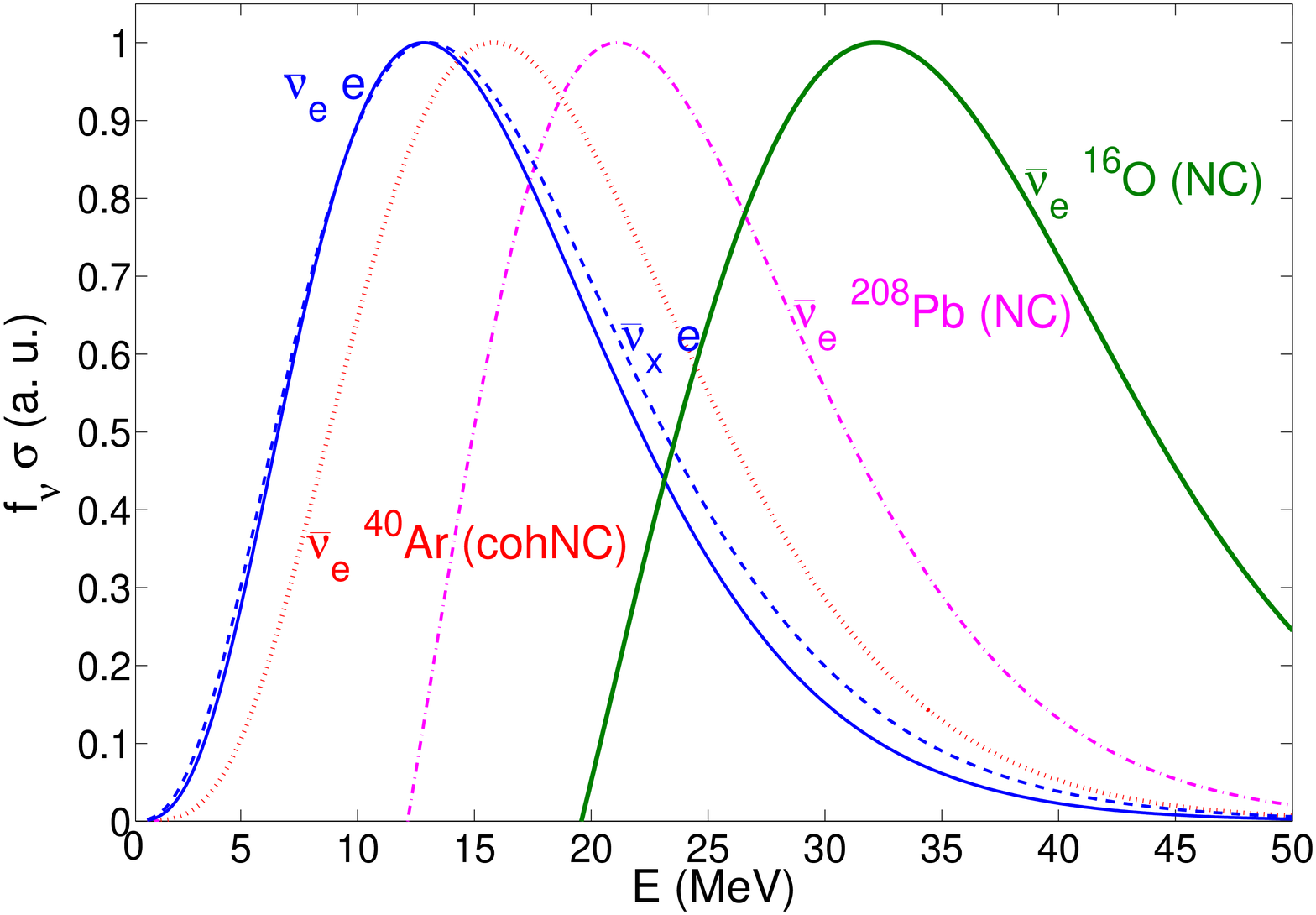}
\vskip12pt
\includegraphics[width=1.\linewidth]{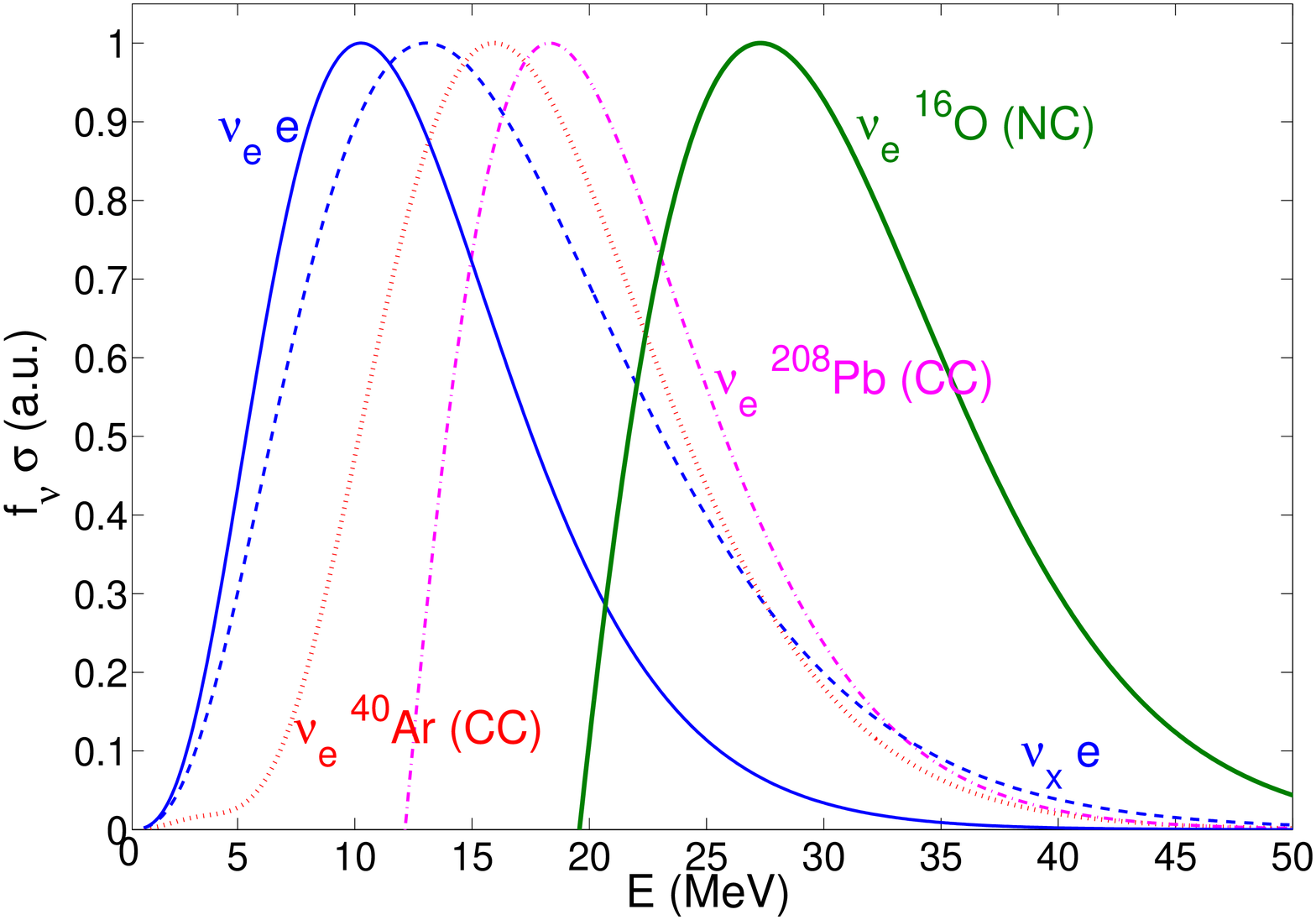}
\caption{\label{fig:rates}Detection rate as a function of neutrino energy
for different reactions, based on our 1016~ms model.  Each curve is normalized to 
its maximum.
{\it Top:} Anti-neutrinos. {\it Bottom:} Neutrinos.}
\end{figure}

The detector counting rate as a function of incoming neutrino energy
is proportional to $f_\nu(E)\sigma(E)$. Based on the numerical
neutrino spectra for the 1016~ms model, we show this quantity in
Fig.~\ref{fig:rates} for our suite of interaction processes,
normalized to the  maximum rate of each process. This plot demonstrates that the
different detection processes probe vastly different parts of the
neutrino spectra.

We now compare the total counting rates for each of our reactions
using as a neutrino spectrum either the numerical output directly
(rate $N_{\rm num}$) compared with the rate inferred if the spectrum
is represented by our fit function (rate $N_{\rm fit}$). In
Table~\ref{tab:table2}, we report the ratio $N_{\rm fit}/N_{\rm
num}$ for neutrinos (top part) and antineutrinos (bottom part). The
agreement between $N_{\rm num}$ and $N_{\rm fit}$ is very good and
typically the error is only a few percent. Even for the oxygen reaction, which
probes the highest-energy tail of all examples, the difference is
only some 10--20\%.  Assuming complete flavor conversions 
 due to neutrino oscillations ($\nu_e \rightarrow \nu_x$, and the same for antineutrinos), 
 we tested that the ratio $N_{\rm fit}/N_{\rm
num}$ differs from the ones reported in Table~\ref{tab:table2} only
on the level of tenths of a percent (results not shown here).
Such a result further confirms the good quality of the 
alpha-fit.

\begin{table}
\caption{Ratio of counting rate based on our
spectral fit and the direct numerical HR spectrum, $N_{\rm fit}/N_{\rm
num}$.
\label{tab:table2}
}
\begin{ruledtabular}
\begin{tabular}{llllll}
$t [\mathrm{ms}]  $ & $\nu_e {\rm ^{40}Ar}$& $\nu_e {\rm ^{208}Pb}$  &  $\nu_e e$  & $\nu_x e$ &
 $\nu_e {\rm ^{16}O}$ \\
  & CC & CC  & & & NC\\
 \hline
261 & 1.01 & 1.02 & 1.00 & 1.00 & 1.01\\
1016 & 1.02 & 1.06 & 1.01 & 1.00 & 0.91\\
1991 & 1.02 & 1.07 & 1.01 & 1.00 & 0.90\\
\hline\hline
\strut\\
\hline\hline
$t [\mathrm{ms}]   $ & $\bar{\nu}_e {\rm ^{40}Ar}$ & $\bar{\nu}_e {\rm ^{208}Pb}$  &  $\bar{\nu}_e e$  & $\bar{\nu}_x e$ &
 $\bar{\nu}_e {\rm ^{16}O}$ \\
 & NC & NC  & & & NC\\
 \hline
261 & 1.00 & 1.02 & 1.00 & 1.00 & 1.05\\
1016 & 1.01 & 1.03 & 1.00 & 1.00 & 1.16\\
1991 & 1.00 & 1.03 & 1.00 & 1.00 & 1.18\\
\end{tabular}
\end{ruledtabular}
\end{table}

\section{Conclusions}                          \label{sec:conclusions}

 In this paper, we have investigated whether the simple
  analytic fit of Eq.~(\ref{eq:alphafit}), based on the first two
  energy moments of the numerical distribution, is sufficiently
  accurate to predict the detector response for target nuclei with
  relatively high threshold energies.  In order to answer this
  question reliably, we have produced high-resolution (HR) neutrino
spectra at several postbounce times of a spherically symmetric SN
simulation.    In addition, these spectra have allowed us to check for
the first time whether the coarser standard resolution (SR) of typical
supernova models (which is dictated by strong CPU-time limitations)
already allows robust signal predictions.

We have compared SR and HR spectra and fitted both with the analytic fit
of Eq.~(\ref{eq:alphafit}), based on the first two energy moments of
the numerical distribution.  These fitted spectra account well for the
detection rates in SN neutrino detectors with vastly different target
nuclei.  We conclude that  for the purpose of signal
  forecast in different detectors reasonably good accuracy can be
  achieved by considering the lowest two energy moments of numerically
  produced spectra. Moreover, we have verified that a modest standard
  energy resolution as affordable in typical supernova simulations
  already gives energy moments surprisingly close to the HR case.  This
insight simplifies parametric studies of detection forecasts for the
neutrino signal from the next nearby SN because it is sufficient to
represent the non-equilibrium spectra by a simple three-parameter
quasi-thermal fit.

\section*{Acknowledgments}
We acknowledge partial support by the Deutsche
Forschungsgemeinschaft under grant TR-7 ``Gravitational Wave
Astronomy'' and the Cluster of Excellence EXC-153 ``Origin and
Structure of the Universe'' and by the European Union Initial
Training Network Invisibles PITN-GA-2011-289442.
I.T.~acknowledges support by the Alexander von Humboldt Foundation.


\end{document}